\documentclass{iau} 
\usepackage{graphicx}

\title[S 346.~~Common envelope evolution of massive stars] 
{Common Envelope Evolution of Massive Stars}

\author[Paul M. Ricker et al.]   
{Paul M. Ricker$^1$, Frank X. Timmes$^2$, Ronald E. Taam$^3$, \and Ronald F. Webbink$^1$}

\affiliation{$^1$Department of Astronomy, University of Illinois, \\ 1002 W. Green St.,
Urbana, IL 61801 USA \\ email: {\tt pmricker@illinois.edu, rwebbink@illinois.edu} \\[\affilskip]
$^2$School of Earth and Space Exploration, \\
Arizona State University, Tempe, AZ 85287-1404 USA \\ email: {\tt Francis.Timmes@asu.edu} \\[\affilskip]
$^3$Department of Physics and Astronomy, Northwestern University, \\ 2145 Sheridan Road, Evanston, IL 60208 USA \\
email: {\tt r-taam@northwestern.edu}}

\pubyear{2018}
\volume{346}  
\jname{High Mass X-ray Binaries: illuminating the passage from massive binaries to merging compact objects}
\editors{A.C. Editor, B.D. Editor \& C.E. Editor, eds.}
\begin{document}

\maketitle

\begin{abstract}
The discovery via gravitational waves of binary black hole systems with total masses greater than $60M_\odot$ has raised
interesting questions for stellar evolution theory. Among the most promising formation channels for these systems is
one involving a common envelope binary containing a low metallicity, core helium burning star with mass $\sim 80-90M_\odot$
and a black hole with mass $\sim 30-40M_\odot$. For this channel to be viable, the common envelope binary must eject more
than half the giant star's mass and reduce its orbital separation by as much as a factor of 80. We discuss issues
faced in numerically simulating the common envelope evolution of such systems and present a 3D AMR simulation of
the dynamical inspiral of a low-metallicity red supergiant with a massive black hole companion.
\end{abstract}

\firstsection 
\section{Introduction}

Of the five statistically significant gravitational wave detections to date, three have involved binary black
hole mergers with at least one component having a mass greater than $30M_\odot$ \cite[(Abbott et al. 2016, 2017a, b)]{Abbott2016,Abbott2017a,Abbott2017b}. While these objects were not unexpected \cite[(Lipunov, Prostnov, \& Prokhorov 1997)]{Lipunov1997}, at metallicities close to solar very massive stars generally lose too much matter to winds to leave behind such massive black holes \cite[(Belczynski et al. 2010)]{Belczynski2010}.
There are, nevertheless, a number of different evolutionary channels through which pairs of massive binary black holes might
form. The two most-studied categories of models are isolated binary channels
and dynamical channels that bring together
black holes formed separately \cite[(see Tutukov \& Cherepashchuk 2017 for a recent review)]{Tutukov2017}.

Here we examine the common envelope (CE) mass-transfer phase that is an ingredient of the `classical'
isolated compact binary formation channel \cite[(Tutukov \& Yungelson 1973; van den Heuvel \& De Loore 1973)]{Tutukov1973,vandenHeuvel1973}. We consider in particular the `typical' model discussed by
\cite{Belczynski16}. In this model, a low-metallicity binary with zero-age main sequence masses of about $96M_\odot$ and
$60M_\odot$ first undergoes a mass-transfer phase that inverts the mass ratio but does not produce much orbital
shrinkage. After the first black hole forms through direct collapse and the second star reaches core helium burning,
the binary undergoes a
CE phase that shrinks the orbit by a factor of 80 or more and ejects half of the donor star's mass. This level of
orbital shrinkage, which is necessary in order to bring the final stellar remnants close enough to merge due to
gravitational radiation emission within a Hubble time, requires a high envelope ejection efficiency and/or sources
of energy beyond the potential energy of the binary orbit \cite[(Kruckow et al. 2016)]{Kruckow16}. Our objective
is to determine if this orbital shrinkage is a realistic prediction. In the process we will
examine a mass range not heretofore considered in three-dimensional CE simulations, which have for the most part
focused on low-mass systems \cite[(Ivanova et al. 2013)]{Ivanova13}.

CE simulations involving very massive donor stars offer challenges beyond those required for low-mass systems.
The spatial dynamic range required to simultaneously resolve the helium-burning core of a supergiant and the
ejected envelope can be a factor of ten larger than that required for a low-mass AGB star. The envelope thermal
timescale is comparable to or less than the dynamical timescale, requiring radiative transfer
to be included. Winds may be important during CE mass transfer. Finally, the large Eddington factors in
massive-star envelopes change the behavior of convection there in ways that are just beginning to be understood
\cite[(e.g. Jiang et al. 2018)]{Jiang18}. The simulations described here represent a first attempt to address
the CE problem in this regime; a more complete investigation is underway and will be discussed in a forthcoming paper.

\section{Numerical methods}

We use the adaptive mesh refinement (AMR) code FLASH 4.5 \cite[(Fryxell et al. 2000; Dubey et al. 2008)]{Fryxell00,Dubey08} to simulate common envelope evolution.
Simulations are carried out on an oct-tree mesh via the PARAMESH library \cite[(MacNeice et al. 2000)]{MacNeice00} using $8^3$ zones per block with
the base mesh level formed from $32^3$ blocks. The mesh is refined using the standard second-derivative criterion \cite[(L\"ohner 1987)]{Lohner87}
for blocks with densities above $3\times10^{-11}{\rm \ g\ cm}^{-3}$. The Euler equations are solved using the directionally split
Piecewise Parabolic Method \cite[(PPM; Colella \& Woodward 1984)]{Colella84} with a variant of the Helmholtz equation of state that includes hydrogen and
helium partial ionization via table lookup. The gravitational potential is computed using a direct multigrid solver \cite[(Ricker 2008)]{Ricker08a}
with isolated boundary conditions. The companion star and the core of the donor star are represented using particles. In contrast
to our earlier work \cite[(Ricker \& Taam 2008, 2012)]{Ricker08b,Ricker12}, instead of using particle clouds we use single particles whose gravitational accelerations
are directly added to the mesh accelerations computed by differencing the mesh potential. The particles are treated as corresponding to
uniform-density spheres with radius equal to three times the smallest zone spacing. We include single-group flux-limited radiation diffusion using
Crank-Nicolson integration and the HYPRE linear algebra library with the \cite{Levermore81} flux limiter. Opacities are determined using
the OPAL tables \cite[(Iglesias \& Rogers 1996)]{Iglesias96} for
high temperatures and the \cite{Ferguson05} tables for low temperatures, both with $Z=0.0001$.

To construct initial conditions for FLASH, we evolve the donor star from the zero-age main sequence using MESA \cite[(Paxton et al. 2011, 2013, 2015, 2018)]{Paxton11,Paxton13,Paxton15,Paxton18}. Once the star reaches maximum expansion we remove the envelope density inversion that develops by expanding the outermost layers of the star at constant entropy until hydrostatic equilibrium is reached (this procedure stands in place of a more sophisticated treatment still to be developed). We then
relax the star in the binary potential together with the companion using a 
heavily modified version of the 3D smoothed particle hydrodynamics (SPH) code StarCrash \cite[(Rasio \& Shapiro 1992; Faber \& Rasio 2000)]{Rasio92,Faber00}. The SPH code has been modified to use a
tree solver for gravitation, a variable-timestep leapfrog integrator, the same equation of state used in FLASH, a Lagrangian formulation to correctly
include the effects of variable smoothing length, and the same particle cores used in FLASH. It also includes an explicit radiation diffusion solver;
an implicit solver based on the method of \cite{Whitehouse05} is under development. To initialize the SPH code from the MESA model, we employ an
approach similar to that suggested by \cite{Ohlmann17}, solving a modified Lane-Emden equation with density and entropy matched to the MESA model at
the numerical core radius.

Further details of the numerical methods used and tests of the initialization procedure will be presented in a
forthcoming paper.

\section{Results}

We conducted CE simulations of a binary system containing an $82.1M_\odot$ red supergiant (RSG) and a $35M_\odot$ black hole. The RSG had a metallicity $Z= 0.0002$ and was evolved from an $88M_\odot$ zero-age mass until it reached maximum expansion at a radius of $2891R_\odot$. After flattening the outer envelope to remove the density inversion, the star's radius increased by 11.7\%. Two runs with somewhat different resolutions were conducted:
a nonradiative run with box size of 375~AU, minimum zone spacing $78.6R_\odot$, and core mass $63.7M_\odot$; and a
run including radiation diffusion having a box size of 274~AU, minimum zone spacing $115R_\odot$, and core mass
$63.7M_\odot$. These resolutions represent a minimum physically reasonable value for this donor star, as its
envelope binding energy measured relative to the surface increases sharply in magnitude inside a radius of $\sim 200R_\odot$. Removing the envelope becomes progressively more difficult at smaller radii.

\begin{figure}[hbt]
\begin{center}
 \includegraphics[height=1.95in]{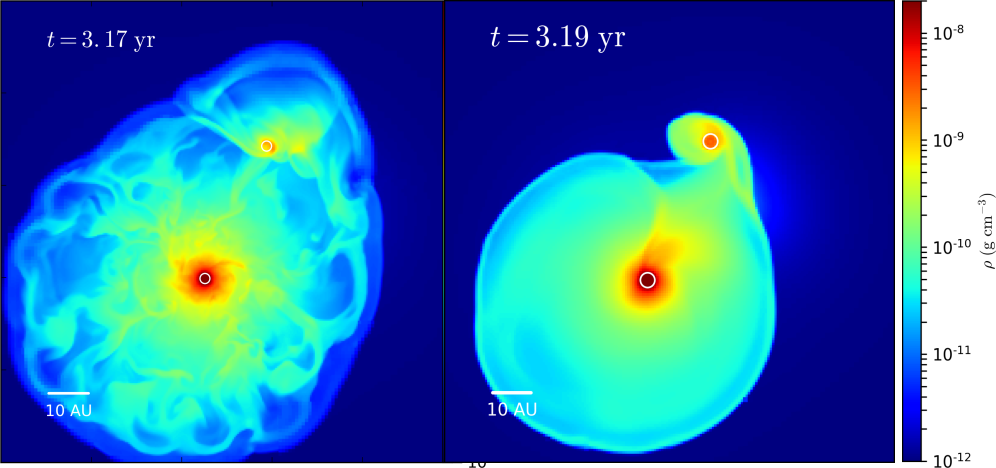} 
 \caption{\textit{Left:} Gas density in the orbital plane in the nonradiative run after 3.17~yr. Circles represent
 the positions and sizes of the donor core and the companion. \textit{Right:} Same, but for the radiative run after
 3.19~yr.}
   \label{fig1}
\end{center}
\end{figure}

A comparison of the nonradiative and radiative runs at an early stage is shown in Fig.\,\ref{fig1}. Large convective
eddies seen in the nonradiative run are absent in the radiative run. This occurs because the thermal readjustment
timescale of the envelope is less than or equal to the dynamical timescale. The envelope is convectively unstable,
but because of the short thermal timescale the heat flux implied by the temperature gradient should be carried
partly by radiative diffusion. In the nonradiative run this is not possible, so convection is far more vigorous
than it should be. This behavior is not seen in CE simulations involving low-mass giants because the thermal
timescales of such stars are much longer.

\begin{figure}[hbt]
\begin{center}
 \includegraphics[height=1.95in]{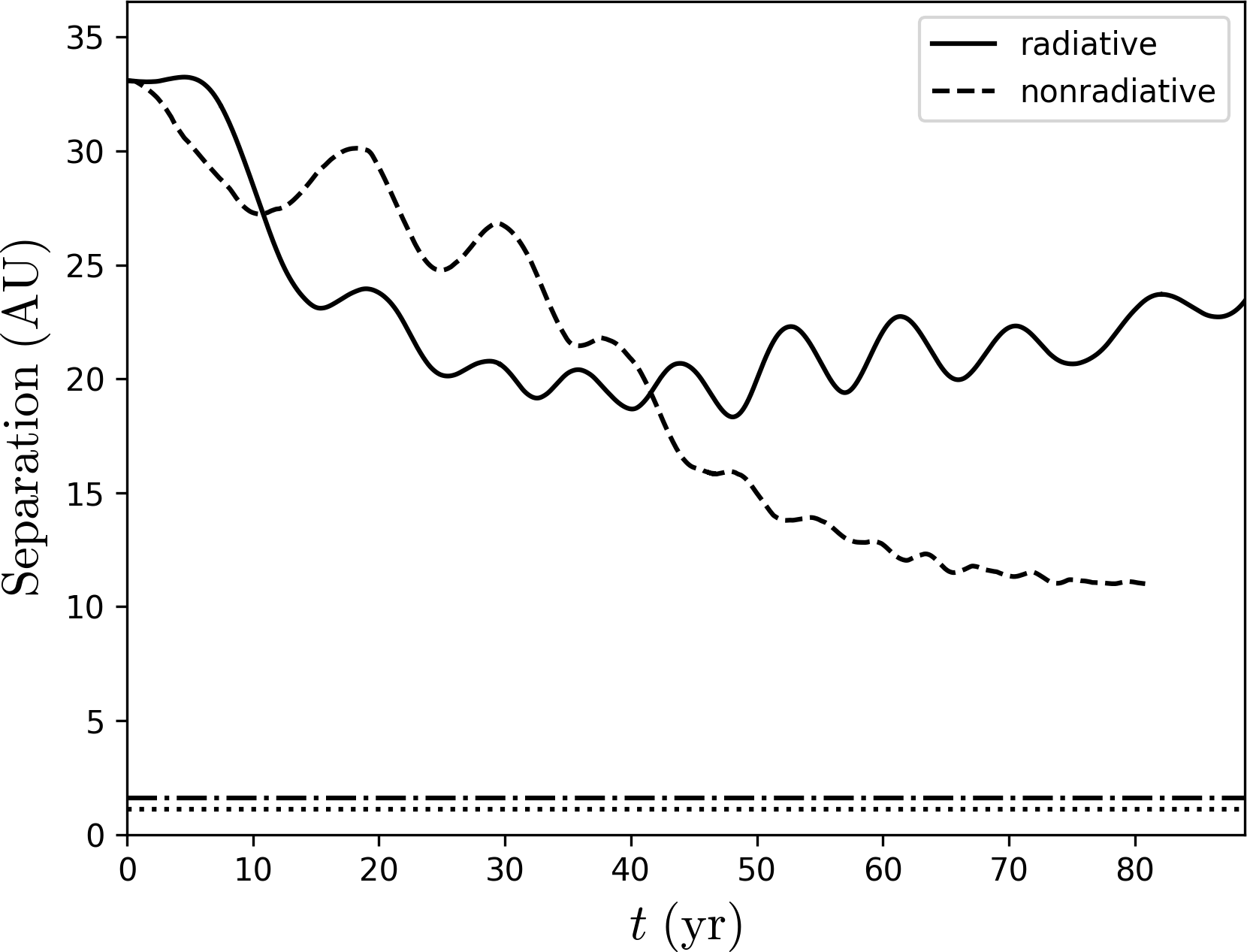} \ 
 \includegraphics[height=1.95in]{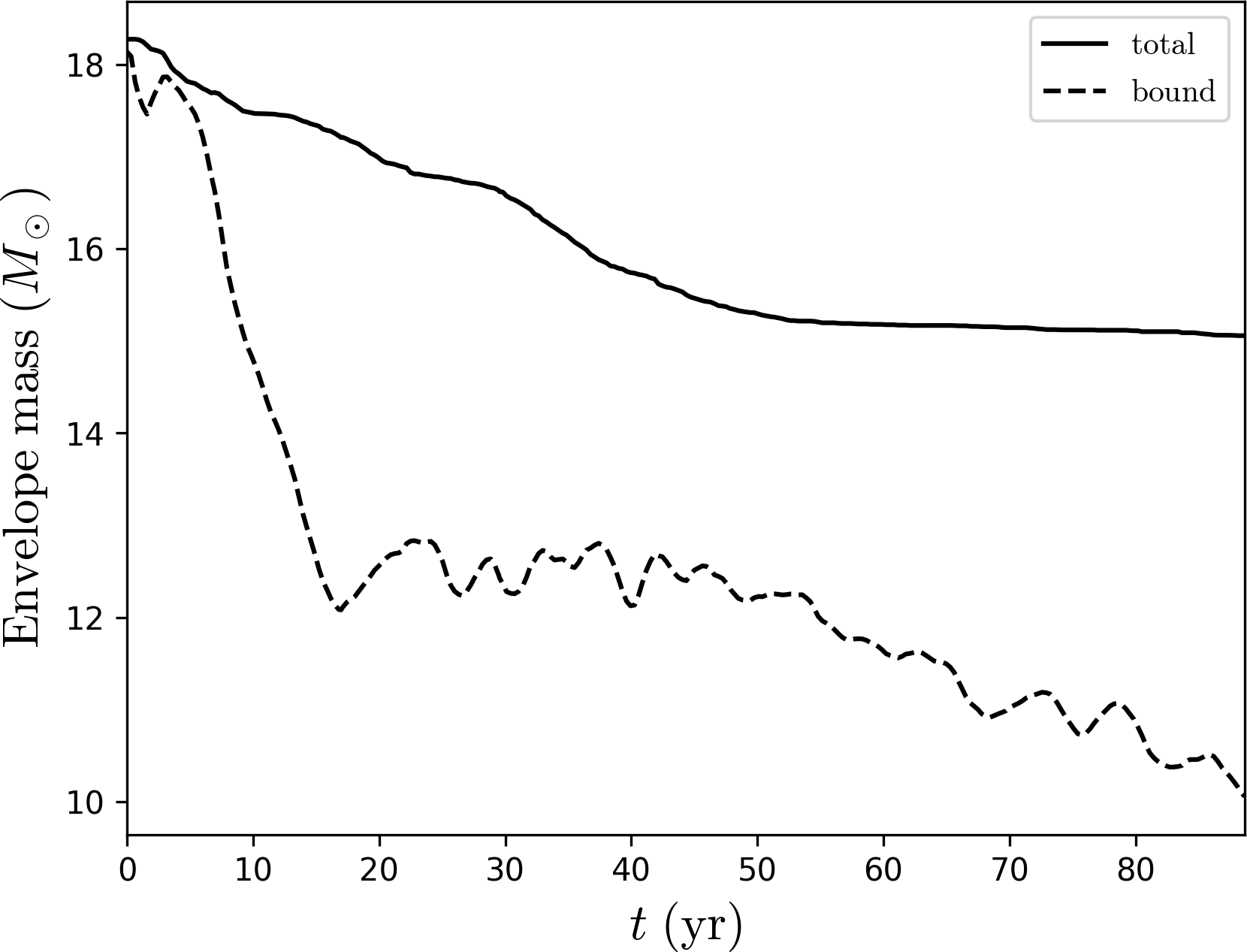} 
 \caption{\textit{Left:} Orbital separation vs.\ time in the nonradiative and radiative runs. Dotted and
 dash-dotted lines represent the core radius in the nonradiative and radiative runs, respectively. \textit{Right:} Total and bound gas mass vs.\ time in the radiative run.}
   \label{fig2}
\end{center}
\end{figure}

In Fig.\,\ref{fig2} we show the orbital separation and gravitationally bound gas mass versus time in the radiative run. Initially the orbital separation and period are 33.1~AU and 17.6~yr, respectively. At first the stars
inspiral by about a factor of two and demonstrate orbital circularization as in previous
simulations in the literature. However, after about 40~yr of evolution the orbital eccentricity increases briefly,
and the orbit widens again to 23.4~AU. In the nonradiative case the orbit shrinks to a separation of about 11~AU
before apparently stabilizing at a constant value without passing through a phase of increasing eccentricity.

For the radiative run, the initial plunge corresponds to a reduction of about $6M_\odot$
in the bound mass, or about 1/3 of the gas initially on the grid. This unbinding appears to stall for about
20~yr before beginning again. By the end of the run (88.7~yr), $8.3M_\odot$ have been unbound, with the trend
toward continued unbinding. The total mass (bound plus unbound) declines until about 55~yr before becoming nearly
constant. Since the hydrodynamical method and AMR library explicitly conserve mass, this means that mass outflow
from the grid has stopped at least temporarily.

\section{Discussion}

The lack of sufficient orbital shrinkage observed in our simulations is in line with results from low-mass CE
simulations \cite[(e.g., Ricker \& Taam 2008, 2012; Passy et al. 2012; Ohlmann et al. 2016; Iaconi et al. 2017)]{Ricker08b,Ricker12,Passy12,Ohlmann16,Iaconi17}. Energy release due to hydrogen and helium recombination
has been proposed as an aid to envelope ejection \cite[(e.g., Webbink 2008)]{Webbink08} and does appear to help in double white dwarf formation \cite[(Nandez, Ivanova, \& Lombardi 2015)]{Nandez15}, though there is some dispute over whether this energy
is simply radiated away \cite[(Soker, Grichener, \& Sabach 2018; Ivanova 2018)]{Soker18,Ivanova18}.
For massive stars the available recombination energy is generally less than for low-mass stars at maximum expansion; the amount depends on the core definition but is not sensitive to metallicity \cite[(Kruckow et al. 2016)]{Kruckow16}. Since our radiative simulation includes (for the first time in 3D CE simulation work) both
the effects of partial ionization on the equation of state and radiative transfer effects, we are able to
address directly the question of whether recombination energy helps in envelope removal for very massive stars.
The initial results presented here suggest that it does not. In a separate project we are also investigating the
efficacy of recombination energy in low-mass envelopes (Zhu et al., in prep.).

An alternative source of energy, which we have yet to consider, is accretion onto the compact object, leading to
jets or accretion disk winds
\cite[(Soker 2004, 2017; Chamandy et al. 2018; L\'opez-C\'amara, De Colle, \& Moreno M{\'e}ndez 2018)]{Soker04,Soker17,Chamandy18,Lopez-Camara18}. However, results from 1D modeling suggest that accretion disk
formation may be a transitory phase in the very massive stars we are considering here \cite[(Merguia-Berthier et al. 2017)]{Merguia-Berthier17}.

The fact that the orbit widening, outflow cessation, and resumption of unbinding all begin at roughly the same
time ($\sim 40-50{\rm \ yr}$) suggests that they may be connected. The ratio of the
kinetic plus thermal energy to the gravitational energy for the unbound material remaining on the grid
is roughly constant at a
value of about 1.5 until 40~yr, after which time it begins to steadily increase, reaching a value of 4 after
88.7~yr. The computational volume may not be large
enough to contain all of the matter whose gravitational influence matters. An additional consideration is the fact
that at the end of the radiative run the orbit appears to be re-circularizing as the bound fraction of the mass remaining on the grid continues to decrease. It is possible that longer-term evolution of the system studied here
may reveal a new phase of orbital shrinkage. We will address this question in future work.

\medskip

PMR acknowledges support from the National Science Foundation under grant AST 14-13367, as well as the
hospitality of the Academia Sinica Institute for Astronomy and Astrophysics during a sabbatical visit.
Portions of this work were completed at the Kavli Institute for Theoretical Physics, where it was supported
in part by the National Science Foundation under grant NSF PHY-1125915. FLASH was developed and is maintained largely by the DOE-supported Flash Center for Computational Science at the University of Chicago. Simulations
were carried out using XSEDE resources at the Texas Advanced Computing Center under allocation TG-AST040034N.

\begin{discussion}

\discuss{van den Heuvel}{We heard in this meeting of the highly obscured B[e] HMXBs like Chaty's INTEGRAL source with an 80 day orbit with an enormous amount of dust around. It does not seem to be spiralling in, which may be a confirmation of your simulations.}

\discuss{Ricker}{Possibly. But this is a low-metallicity system, so the effects of dust opacity should be much
less than for the INTEGRAL source.}

\discuss{Sander}{The density inversion in massive stars is also something that bothers us when analyzing stars with stellar atmospheres. However, we have issues with temperatures for certain massive stars that are lower than we expect. There are also theoretical works (e.g. Gr{\"a}fener et al., Sanyal et al.) pointing towards an inflated envelope. Couldn't that be something that might eventually even help to eject the CE?}

\discuss{Ricker}{Some inflation is already present here due to our treatment of the density inversion in the MESA model. Further expansion might help increase tidal drag, but it would also make it easier for recombination energy to escape, so it's not clear what the net effect would be.} 

\end{discussion}

\end{document}